\documentclass[12pt]{article}
 \usepackage{amsmath}

\title{Magnetic Catalysis and Quantum Hall Ferromagnetism in Weakly Coupled Graphene}
\author{Gordon W.~Semenoff and Fei Zhou\\~~\\Department of Physics and Astronomy, University of British Columbia,\\
Vancouver, British Columbia, Canada V6T 1Z1}

\begin{document}
\maketitle

\begin{abstract}
We study the realization in a model of graphene of the phenomenon whereby the tendency
of gauge-field mediated interactions to break chiral symmetry spontaneously
is greatly enhanced in an external magnetic field. We prove that, in the weak coupling
limit, and where the electron-electron interaction satisfies certain mild conditions,
the ground state of charge neutral graphene in an external magnetic field
is a quantum Hall ferromagnet which spontaneously breaks the emergent
U(4) symmetry to U(2)XU(2).
 We argue that, due to a residual CP symmetry, the quantum Hall ferromagnet order parameter is given exactly by the
leading order in perturbation theory. On the other hand, the chiral condensate which is the order
parameter for chiral symmetry breaking generically obtains contributions at all orders.  We
compute the leading correction to the chiral condensate.  We argue that the ensuing fermion
spectrum resembles that of massive fermions with a vanishing U(4)-valued chemical potential.
We discuss the realization of parity and charge conjugation symmetries and
argue that, in the context of our model,
the charge neutral quantum Hall state in graphene is a bulk insulator, with vanishing longitudinal
conductivity  due to a charge gap and
Hall conductivity vanishing due to a residual discrete particle-hole symmetry.
\end{abstract}

\section{Introduction and Summary}

Graphene is a 2-dimensional honeycomb array of carbon atoms
whose electronic properties at energies below a few electron volts
are modeled by  emergent
relativistic Dirac fermions with   $U(4)$
chiral symmetry~\cite{Semenoff:1984dq}-\cite{Gusynin:2005pk}.   Though all indications to date
are that clean, suspended graphene is indeed a gapless semi-metal, the idea that
Coulomb and other interactions  could potentially gap the electron spectrum
by breaking the $U(4)$ symmetry
has inspired considerable discussion \cite{cat5}-\cite{chiral7} and some numerical
work \cite{Armour:2009vj}-\cite{drut2010B}.
Moreover, recent experiments
show plenty of evidence~\cite{zhang}-\cite{ghahari2011} that some form of symmetry
breaking does occur in graphene at sufficiently low temperatures and in
strong magnetic fields. Understanding the mechanism for this magnetic field driven
symmetry breaking has become an important problem.

It has been known for a long time that
an external magnetic field enhances the tendency of interactions to break chiral
symmetry and generate a mass gap in 2+1 dimensional quantum field theories with massless fermions~\cite{cat0}.
This idea, which is called ``magnetic catalysis'',  has already been applied to graphene
and graphene-like systems~\cite{cat1}-\cite{cat4},\cite{cat5}\cite{cat6}.
The mechanism of magnetic catalysis is simple to understand \cite{cat6}\cite{ssw}.
A magnetic field flattens the electron energy bands by quenching
the kinetic energy to form Landau levels. In a state where
a given Landau level must be partially filled, one must therefore adjust the way in which the levels are filled to
minimize the potential energy, as
all configurations have approximately the same kinetic energy. Then, even a very weak interaction which favors an asymmetric population of the single particle states in the band will break symmetry.

An alternative, but closely related scenario for breaking the $U(4)$ symmetry of a Landau level is
called quantum Hall ferromagnetism~\cite{qhf}. This idea is inherited from  multi-layer
quantum Hall systems made from conventional semiconductors where it explains the appearance of ferromagnetic Hall states \cite{qhf1}\cite{qhf2}.
The quantum Hall ferromagnet is designed to minimize the Coulomb energy of a single Landau level by making the electronic
wave-function as anti-symmetric in the electron positions as possible, since this  gives the electrons as much
spatial separation
as possible and minimizes the repulsive Coulomb potential energy.
For the lowest Landau level of graphene\footnote{We shall refer to the Landau level which
apprears at the apex of the graphene Dirac cones as the ``lowest Landau level''.  It is also called the ``charge neutral point'' Landau level.}, this is accomplished by making the $U(4)$ indices of electrons occupying the Landau level
as symmetric as possible which for some commensurate fillings; quarter-filling, half-filling or three-quarters-filling of the level
is a $U(4)$ ferromagnetic state that breaks the $U(4)$ symmetry spontaneously. If mixing of Landau levels is ignored and
the dynamics restricted to the charge
neutral point
Landau level,  the quantum Hall Ferromagnet is
an exact eigenstate of the charge density operator and it can minimize the energy due to
a density-density interaction like the
Coulomb interaction.

Quantum Hall ferromagnetism is sometimes viewed as a competitor of magnetic catalysis.
The main distinguishing feature of the two are
the order parameters which for quantum Hall ferromagnetism is a Hermitian matrix-valued U(4) electron density \begin{equation}\rho^{\sigma\sigma'}=\left<\frac{1}{2}
\left[\Psi_\eta^{\sigma'\dagger}(x),\Psi_\eta^\sigma(x)\right]\right>
 \label{rho}
\end{equation}
whereas, for magnetic catalysis, it is an electron mass operator condensate\footnote{For a discussion of the
most general form of fermion bilinear condensates in graphene, see Ref.~\cite{chamon} and also \cite{tanaka1}-\cite{herbut2007}.}
\begin{equation}\label{sigma}
\Sigma^{\sigma\sigma'}=\left<\frac{1}{2}\left[\bar\Psi_\eta^{\sigma'} (x),\Psi_\eta^\sigma(x)\right]\right>
\end{equation}
In the above equations, $\Psi_\eta^\sigma(x)$ is the second-quantized electron field operator (see (\ref{fullhamiltonian}),
(\ref{fieldanticommutator}), (\ref{diracmatrices}) below).
 The repeated indices $\eta$ in (\ref{rho}) and (\ref{sigma}) are summed over the pseudo-spin
components, $\eta=1,2$ and $\sigma$ and $\sigma'$
are $U(4)$ (spin and valley) indices with values $\sigma,\sigma'=1,...,4$.  Under a $U(4)$ transformation implemented with $4\times4$ unitary
matrix $U$, both    $\rho^{\sigma\sigma'}$ and $
\Sigma^{\sigma\sigma'}$ are Hermitian matrices which transform by conjugation,
$
\rho \to U\rho U^\dagger ~~,~~ \Sigma \to U\Sigma U^\dagger
$. 
The Dirac conjugate is defined as $\bar\Psi^\sigma=\Psi^{\sigma\dagger}\gamma^0$
and we will use a convention (in (\ref{hamiltonian}) below) for Dirac matrices where
$\gamma^0_{\eta\eta'}={\rm diag}(1,-1)$.
The  commutator ordering in (\ref{rho}) subtracts a constant vacuum charge density.  It is important
for giving the charge density operator the correct charge conjugation (particle-hole) symmetry. The  constant
that is subtracted is the
contribution to the charge density of the positive Carbon ions in graphene which is such that the
particle-hole symmetric point in the electron spectrum is also charge neutral.
This normal
ordering is also helpful for defining the operator in
(\ref{sigma}), as the individual terms in its computation are divergent and the commutator cancels
some of the   most severely divergent parts.

However, it was observed in Ref.~\cite{cat6}, and we will also make this point in the following,
that in a $U(4)$ symmetry breaking phase in a magnetic field, the  order parameters (\ref{rho}) and (\ref{sigma})
can both be non-zero.
In this Paper, we will study this issue using
Hamiltonian techniques and perturbation theory.  We will concentrate on the case of charge neutral graphene in
a magnetic field and at zero temperature.
 In neutral graphene, when the $U(4)$ symmetry is unbroken, the four-fold degenerate
Landau level which resides at the charge neutral Dirac point
should be half-filled.  It is the fact that this level is half-filled which is responsible for the anomalous
integer Hall effect that was observed in early experiments \cite{Novoselov:2005kj}\cite{halleffect} -- the charge neutral point is half way up a Hall conductivity step,
rather than being on a Hall plateau, as it would
be in a conventional integer quantum Hall system.

Symmetry breaking would have this four-fold degenerate Landau level splitting into  levels
of lower degeneracy. The relevant splitting results in either two
 2-fold degenerate levels or four nondegenerate levels, half of which have energy lower than the Dirac point and the
other half have energy above the Dirac point.  In the charge neutral case, the spectrum is then gapped. At zero temperature  the levels below the Dirac point are completely filled and those above are completely empty.
The result is a new Hall plateau at the charge neutral point.  This plateau has been observed in clean suspended graphene at low temperatures and high magnetic fields~\cite{zhang}-\cite{gapopening}. As we shall argue, discrete symmetries imply that the bulk Hall conductivity of this state should vanish, i.e., the charge neutral Hall plateau
is a quantum Hall insulator with bulk conductivities vanishing, $\sigma_{xy}=0$ and $\sigma_{xx}=0$.
Of course currents can be carried by edge states
which could exist. It is also known that the occurrence of edge states and their detailed properties depend
 on the $U(4)$ orientation of the condensate which we shall not address here.

Our study will be limited to the case of weak
interactions of the specific charge density-charge density form given in the Hamiltonian (\ref{fullhamiltonian}) below,
an important example of which is
the Coulomb interaction. We will ignore impurities as well as explicit $U(4)$ symmetry breaking, which would
be small but present in graphene due to lattice
effects and Zeeman and spin-orbit coupling.
For simplicity, we will concentrate on the charge neutral Hall state
and focus on  bulk properties of graphene.  We will not discuss edge currents which are generally important for the
phenomenology of quantum Hall states.
We shall hereafter refer to ``graphene'' as the U(4) symmetric
continuum quantum field theory with Hamiltonian  (\ref{fullhamiltonian}).

The weak coupling limit has the advantage that we can definitely identify the
correct ground state.  Our results will show that, if interactions
are repulsive and are weak enough, in the context of our model, charge neutral graphene in a
magnetic field indeed experiences spontaneous breaking of the $U(4)$ symmetry to $U(2)\times U(2) $.
Furthermore, a particular form of the quantum Hall ferromagnet
state proposed in Ref.~\cite{qhf}, with perturbative corrections from Landau-level mixing,
is the preferred ground state. The condensate
contains a non-zero expectation value of a density and a mass operator,
which in the leading order have identical magnitudes, but which
have significant differences when interactions are taken into account.

In the leading order, the condensates $\rho^{\sigma\sigma'}$ and $\Sigma^{\sigma\sigma'}$ have equal
magnitudes.  This
is a result of the fact that, in a magnetic field, the single-particle wave-function of an electron in the Landau level
which resides at the Dirac point, i.e.~the zero mode Dirac spinor, has only one or the other pseudospin components
nonzero, not both. This is the upshot of the index theorem applied to the continuum single particle
Dirac Hamiltonian \cite{Semenoff:1984dq}.\footnote{It also implies that a given zero mode wave-function alway has
support on only one of the two triangular sublattices of the hexagonal graphene lattice.  The zero modes in the two valleys
live on opposite sublattices.} This means that zero modes influence the matrix \begin{equation}\label{condensate55}
\left<\frac{1}{2}\left[\Psi^{\sigma'\dagger}_\zeta(x),\Psi^\sigma_\eta(x)\right]\right>=
\frac{1}{2}\left[\begin{matrix} \rho^{\sigma\sigma'}&0\cr0&\rho^{\sigma\sigma'}\cr \end{matrix}\right]_{\eta\zeta} +
\frac{1}{2}\left[\begin{matrix} \Sigma^{\sigma\sigma'}&0\cr0&-\Sigma^{\sigma\sigma'}\cr \end{matrix}\right]_{\eta\zeta}\end{equation}
(where $\eta,\zeta$ are pseudo-spin labels), only in one diagonal component, either the $\eta$-$\zeta$=1-1 component or the 2-2 component,
for definiteness, assume it is the 1-1 component. (Off-diagonal
components in the indices $\eta,\zeta$ would violate rotation symmetry and must therefore vanish in rotationally invariant states.
We  assume that the condensate preserves translation invariance, in that $\rho^{\sigma\sigma'}$ and $\Sigma^{\sigma\sigma'}$ are independent of position. We are also assuming that the $U(1)$ subgroup of $U(4)$ corresponding to overall phase symmetry
is not broken. With these assumptions, Eq.~(\ref{condensate55}) then gives a complete description of the possible bilinear (dimension 2) electron condensates.
If only the 1-1 component can have an expectation
value, the mass and density must be tuned so that $\rho^{\sigma\sigma'}=\Sigma^{\sigma\sigma'}$.
We shall find that this is the case to the leading (zeroth) order in an expansion in the interaction strength.
Higher orders in interactions de-tune these condensates.  We shall argue that the U(4) charge is given exactly
at the zeroth order, that is, that it is not corrected in perturbation theory.  On the other hand, the chiral
condensate gets corrections, generically at all orders in perturbation theory.   Furthermore, in the relativistic
theory, these corrections can be large as, depending on the nature of the interaction, they can be ultraviolet divergent.

Charge neutral graphene has important discrete symmetries, a combination of which survives in the presence of
a magnetic field and restricts the Hall conductivity.
Given the conventions for Dirac matrices in Eq.~(\ref{diracmatrices})  below,  a particle-hole transformation, $C$,
(or charge conjugation in field theory) is implemented by the replacement
\begin{equation}\label{particlehole}
C:~~\left[\begin{matrix}\Psi^\sigma_1 (\vec x)\cr \Psi^\sigma_2 (\vec x)\cr\end{matrix}\right]~\to~\left[\begin{matrix}0&1\cr 1&0\cr\end{matrix}\right]\left[\begin{matrix}\Psi^{\sigma*}_1 (\vec x)\cr \Psi^{\sigma*}_2 (\vec x)\cr\end{matrix}\right]
= \left[\begin{matrix}\Psi^{\sigma*}_2 (\vec x)\cr \Psi^{\sigma*}_1 (\vec x)\cr\end{matrix}\right]
 \end{equation}
This transformation maps the Dirac equation with external magnetic field $B$ onto the same equation where
$B$ is replaced by $-B$. In addition, it
transforms the condensates as
\begin{equation}
C:~~\rho^{\sigma\sigma'}\to -\rho^{\sigma'\sigma}~~,~~\Sigma^{\sigma\sigma'}\to \Sigma^{\sigma'\sigma}
\end{equation}
Note that both $\rho^{\sigma\sigma'}$ and $\Sigma^{\sigma\sigma'}$ go to their transpose.

Parity, $P$, is the reflection of one coordinate $(x_1,x_2)\to(-x_1,x_2)$ and the Dirac spinor transforms as
\begin{equation}\label{parity}
P:~~\left[\begin{matrix}\Psi^\sigma_1 ( x_1,x_2)\cr \Psi^\sigma_2 ( x_1,x_2)\cr\end{matrix}\right]~\to~\left[\begin{matrix}0&1\cr -1&0\cr\end{matrix}\right]\left[\begin{matrix}\Psi^{\sigma}_1 ( -x_1,x_2)\cr \Psi^{\sigma}_2 ( -x_1,x_2)\cr\end{matrix}\right]
= \left[\begin{matrix}\Psi^{\sigma}_2 ( -x_1,x_2)\cr -\Psi^{\sigma}_1 ( -x_1,x_2)\cr\end{matrix}\right]
 \end{equation}
It also maps the Dirac equation with a constant magnetic field $B$ onto the same equation where
$B$ is replaced by $-B$.  As well,
the condensates transform as
\begin{equation}
P:~~\rho^{\sigma\sigma'}\to \rho^{\sigma\sigma'}~~,~~\Sigma^{\sigma\sigma'}\to- \Sigma^{\sigma\sigma'}
  \end{equation}
We remark that, since they interchange the upper and lower components of the spinor,  $C$
and $P$ both interchange the two triangular sublattices of the honeycomb graphene lattice. We also remark that parity as we have defined
it here is unorthodox in that it does not interchange the $K$ and $K'$ points of graphene, as the transformation defined in the
more conventional way would do.  Ours and the conventional one differ by a $U(4)$ transform (see Ref.~\cite{discretesymmetries} for a detailed discussion of the conventional parity, charge conjugation and
time reversal symmetries).

Since the magnetic field changes sign, neither charge conjugation $C$ in (\ref{particlehole}) nor parity $P$ in (\ref{parity})
are good symmetries when the external magnetic field is applied.  However, we still expect that a combination of the two
transformations, $CP$, is a good symmetry since the combination preserves the sign of the magnetic field.
Intuitively, in a given external field, the cyclotron
orbits of particles and holes would have opposite orientations, so even if particles and holes have the same energies,
particle-hole symmetry is lost.
However, we could recover the symmetry by augmenting it with a re-orientation of the space which reverses
the directions of cyclotron orbits and therefore maps a hole cyclotron orbit onto a particle cyclotron orbit.  This reorientation is parity.  The combined transformation can be a symmetry
of a system in a magnetic field.   Indeed, under the transformation $CP:~~B\to B$ and the Dirac equation with magnetic
field is mapped onto itself.
However, the condensates that we are interested in are not automatically invariant under CP, instead they transform as
\begin{equation}\label{cp55}
CP: ~~\rho^{\sigma\sigma'}\to -\rho^{\sigma'\sigma}~,~\Sigma^{\sigma\sigma'}\to- \Sigma^{\sigma'\sigma}
 \end{equation}
and having a non-zero value for either condensate would signal the potential spontaneous breaking of $CP$ symmetry.
We will be interested in the situation where these condensates are not zero, but where they also do not break the $CP$ symmetry.
We remind and stress to
the reader that this applies to the charge neutral state of graphene, which we refer to as the
charge neutral (or which is commonly referred to as the $\nu=0$) Hall state only.
It does not apply to the other Hall states since they occur at a non-zero
 charge density and non-zero charge density would break C and CP symmetry explicitly.


To emphasize the importance of $CP$ invariance, we note that a non-zero Hall conductivity would break
CP and must vanish if CP is
unbroken.  One way to see this is to note that $CP$ forbids the appearance of a Chern-Simons term
in the effective action for the electromagnetic field which could be obtained by integrating out the electron degrees of freedom
\cite{Niemi:1983rq}-\cite{Redlich:1983kn}, a well-defined
procedure once their spectrum is gapped.  Absence of a Chern-Simons term means that the bulk Hall conductivity of the charge neutral Hall plateau should be zero.  It is indeed observed to be zero in experiments \cite{zhang}\cite{ong1}\cite{ong2} where the $\nu=0$ conductivity plateau observed in
strong magnetic fields has $\sigma_{xy}=0$.

If we require that $CP$ is not spontaneously broken, it places a
strong restriction on the form of the condensates, $\rho^{\sigma\sigma'}$ and $\Sigma^{\sigma\sigma'}$.
One possibility which is clear from (\ref{cp55}) is that they could simply be antisymmetric matrices so that
$CP: ~~\rho^{\sigma\sigma'}\to -\rho^{\sigma'\sigma}=\rho^{\sigma\sigma'}~,~\Sigma^{\sigma\sigma'}\to- \Sigma^{\sigma'\sigma}
=\Sigma^{\sigma\sigma'}$ is indeed $CP$ symmetric.  In fact, due to $U(4)$ symmetry, this is the most general constraint on a $CP$-symmetric condensate:  there exists
a basis where both $\rho$ and $\Sigma$ are Hermitian antisymmetric matrices.

We could also use a $U(4)$ transformation to choose a different basis where  $\rho^{\sigma\sigma'}$ is diagonal, rather than being anti-symmetric.
The eigenvalues of an anti-symmetric Hermitian matrix
are real and come in positive-negative pairs, so the most general form of diagonal $\rho$
is  $\rho^{\sigma\sigma'}= ~{\rm diag}\left(\rho_1,\rho_2,-\rho_1,-\rho_2\right)$.
Generally, $\Sigma^{\sigma\sigma'}$ might not commute with $\rho^{\sigma
\sigma'}$ and therefore, it might not be diagonalizable at the same time.
However, in the basis where $\rho$ is diagonal, $\Sigma^{\sigma\sigma'}$ is a unitary transform of an imaginary anti-symmetric $4\times4$
matrix, which is specified by six real parameters.
The upshot of this discussion is that the possible $CP$-invariant
condensates are described by 8 real parameters, which can be taken to be the two independent eigenvalues of $\rho$ and
the six entries in the upper triangle of
 $\Sigma^{\sigma\sigma'}$ in the basis where it is anti-symmetric.

When $\rho^{\sigma\sigma'}$ has the form
$\rho^{\sigma\sigma'}~=~ {\rm diag}(\rho_1,\rho_2,-\rho_1,-\rho_2)$, if $\rho_1\neq\rho_2$, this
condensate would by itself break the $U(4)$ symmetry to the subgroup which commutes with $\rho$, the maximal torus, $U(1)^4$.
In the special case where $\rho_1 = \rho_2 = \rho$, and $\rho^{\sigma\sigma'}={\rm diag}(\rho,\rho,-\rho,-\rho)$
the residual symmetry is greater, the symmetry breaking pattern being $U(4)\to U(2)\times U(2)$.
This is the largest
residual symmetry that is compatible with a non-zero condensate and $CP$ symmetry. A non-zero $\Sigma^{\sigma\sigma'}$ could break some or all of the remaining
symmetry.   What we will consider in the following is the scenario  where this does not
happen\footnote{We note here that
a simple generalization of the Vafa-Witten theorem \cite{Vafa:1983tf}, which  normally applies to vector-coupled
relativistic gauge theories, to the present model of graphene with a Coulomb interaction implies that
CP breaking would not occur.} and the system has the
maximal residual symmetry $U(2)\times U(2)$ (we will later demonstrate that it is indeed
the relevant one), i.e. $\rho^{\sigma\sigma'}$ has the form
\begin{equation}\label{rho55}\rho^{\sigma\sigma'}= {\rm diag}(\rho,\rho,-\rho,-\rho)\end{equation}
and then, to preserve the same symmetries, $\Sigma^{\sigma\sigma'}$ must be of the form
\begin{equation}\label{sigma55}\Sigma^{\sigma\sigma'}= {\rm diag}(\Sigma,\Sigma,-\Sigma,-\Sigma)\end{equation}
In summary, given the symmetry breaking pattern, $U(4)\to U(2)\times U(2)$ preserving $CP$, the fermion bilinear condensate is completely characterized
by the two parameters $\rho$ and $\Sigma$ in (\ref{rho55}) and (\ref{sigma55}), respectively.
The tuning that we have
discussed above (following Eq.~(\ref{condensate55})) gives one additional piece of information, that in the very weak coupling limit,
the amplitudes of the two condensates are almost equal, and
$\Sigma\sim\pm\rho$ and therefore $\Sigma^{\sigma\sigma'}\sim\pm \rho^{\sigma\sigma'}$.
In fact, this is just what we shall find for the quantum Hall
ferromagnet state at the lowest order  of perturbation theory.
For the quantum Hall ferromagnet state,
 \begin{align}\label{leadingorderrho}
\rho &= \frac{|B|}{4\pi}
\\  \label{leadingordersigma}
\Sigma &= \frac{B}{4\pi}\left[~1~+~ \Delta_1~+~\ldots~~\right]
\end{align}
We shall argue that the U(4) charge density is exact to all orders in perturbation theory. The correction to the
chiral condensate, $\Delta_1$, is positive and is of linear order in the interaction potential $V(\vec x - \vec y)$.
The remainder, represented by three dots is at least of order two in the interaction potential.

The leading terms in the above two condensates do not depend on the interaction strength, but only
on the magnitude of the constant magnetic field.  Furthermore, the $B$-dependence must be as
shown to give $\rho$ and $\Sigma$ the correct scaling dimensions of inverse length squared.
The dependence of the condensates on the sign of $B$ in  (\ref{leadingorderrho}) and (\ref{leadingordersigma})
is compatible with $C$ and $P$ transformations.
In the basis where they are antisymmetric,
  C maps $\rho^{\sigma\sigma'}$ and $\Sigma^{\sigma\sigma'}$ to their transpose and $\rho^{\sigma\sigma'}$
obtains a minus sign.  In the basis where $\rho^{\sigma\sigma'}$ and $\Sigma^{\sigma\sigma'}$ are diagonal, C must therefore
contain a U(4) transformation which interchanges the positive and negative eigenvalues of both $\rho^{\sigma\sigma'}$ and $\Sigma^{\sigma\sigma'}$.  This flips the sign of diagonal $\rho^{\sigma\sigma'}$ and $\Sigma^{\sigma\sigma'}$. This transformation,
together with the sign flip of $B$ under $C$ makes the diagonal $\rho^{\sigma\sigma'}$ and $\Sigma^{\sigma\sigma'}$
odd and even under $C$, respectively, as they should be.  $P$  maps
$B\to -B$.  This leaves $\rho$  in (\ref{leadingorderrho}) unchanged and flips the sign of $\Sigma$ in (\ref{leadingordersigma}), as expected.

We compute he term $\Delta_1$ in (\ref{leadingordersigma}) in Section 4. It is of linear order the interaction and it is
entirely the result of Landau level mixing.  Furthermore, it is positive and, depending on the interaction,
it can contain an ultraviolet divergence
which could make it anomalously large. This result suggests that the chiral condensate is much larger than the
charge condensate.  In fact, if we attempt to obtain the condensates from a phenomenological description of
fermions where we make the approximation that they are non-interacting and we add terms to $H_0$
with a U(4) chemical potential $ \int d^2x~ \Psi^{\sigma\dagger} \mu_0^{\sigma\sigma'}\Psi^{\sigma'}$
and mass term $\int d^2x~\bar\Psi^\sigma M_0^{\sigma\sigma'}\Psi^{\sigma'}$ where,
$$\mu_0^{\sigma\sigma'}~=~{\rm diag}~\left(\mu_0,\mu_0,-\mu_0,-\mu_0\right)~~,~~M_0^{\sigma\sigma'}~=~{\rm diag}~\left(m_0,m_0,-m_0,-m_0\right)$$ we would find that the best fit has $\mu_0=0$.  The condensates
in the free theory are then
\begin{align}\label{phenomrho}
\rho_0&=\frac{|B|}{4\pi}{\rm sign}(m_0)\\ \Sigma_0 &= \frac{B}{4\pi}{\rm sign}(m_0)\left[ 1
~+~\sum_{n=1}^\infty ~\frac{2}{\sqrt{1+2|B|n/m_0^2}}~\right]
\label{phenomsigma}
\end{align}
which is very similar to (\ref{leadingorderrho}) and (\ref{leadingordersigma}). Note that, in (\ref{phenomsigma}), the second term in $\Sigma_0$ is
formally ultraviolet divergent and it is entirely due to the contribution of higher Landau levels.
\footnote{It has
meaningful interpretation once the divergent part is dealt with by multiplicative renormalization.}
It would thus be reasonable to interpret our result for the condensates in the interacting
theory to be entirely due to mass generation. The mass that is generated breaks the U(4) symmetry
to U(2)$\times$U(2) and it preserves C and P invariance (and the magnetic field further reduces C and P to CP).

The model of graphene that we shall use has four species of relativistic
electrons with $U(4)$ symmetry and an instantaneous interaction
\begin{eqnarray}
H&=&H_0+H_I\nonumber \\&=& \int d^2x~\sum_{\sigma=1}^4\sum_{\eta,\zeta=1}^2
 \Psi^{\sigma\dagger}_\eta(\vec x) h_{\eta\zeta}\Psi^\sigma_\zeta(\vec x)  + \frac{1}{2}
\int d^2 x d^2 y~ \hat\rho(\vec x)V(\vec x-\vec y)\hat\rho(\vec y)
\label{fullhamiltonian}
\end{eqnarray}
 where $H_0$ is the non-interacting Hamiltonian given in the
first term and $H_I$ is the interaction Hamiltonian in the second term.
$\Psi^\sigma_\eta(\vec x) $ is the second
quantized electron field obeying the anti-commutation relation
\begin{equation}\label{fieldanticommutator}
\left\{ \Psi^{ \sigma}_\eta(\vec  x),\Psi^{\sigma'\dagger }_\zeta(\vec  y)\right\} =
\delta^{\sigma\sigma'}\delta_{\eta\zeta}\delta( \vec x- \vec y)
\end{equation}
The single-electron Hamiltonian has the form
\begin{equation}\label{hamiltonian}
h_{\eta\zeta}= \hbar v_F\left[ \begin{matrix} 0 & -iD_x-D_y \cr
-iD_x+D_y & 0 \cr \end{matrix}   \right]_{\eta\zeta}
\end{equation}
Here, the conventions for relativistic Dirac matrices are
\begin{align}
\gamma^0= \left[ \begin{matrix} 1 & 0 \cr 0 & -1 \cr \end{matrix}\right]~,~
\gamma^1= \left[ \begin{matrix} 0  & 1 \cr -1 & 0 \cr \end{matrix}\right]~,~
\gamma^2= \left[ \begin{matrix} 0  & -i \cr -i & 0 \cr \end{matrix}\right]
\label{diracmatrices}
\end{align}

The Hamiltonian (\ref{hamiltonian})  describes electron dynamics at energies near
the Dirac points~\cite{Semenoff:1984dq}.
 We shall use a convention where both graphene valleys
have the same Hamiltonian.  This convention
assigns the upper and lower components of the spinors in each valley to the two graphene
sublattices differently -- the upper/lower component in one valley live on sublattice A/B and the
upper/lower component in the other valley reside on sublattices B/A.
In this way, both valley and both spin states have
the same single-particle Hamiltonian and the emergent $U(4)$ symmetry is explicit.
We will work in units where $\hbar=1$ and the emergent speed of light,
the graphene Fermi velocity, $v_F=1$. The covariant derivative,
$D_i= \partial_i -iA_i$, couples the electrons
to a constant external magnetic field $B$, and we could use either the
symmetric gauge $\vec A(\vec x) = \frac{B}{2}\left(- y,x\right)$ or the
Landau gauge $\vec A(\vec x) = B\left( 0,x\right)$, for example.
The  charge density is given by the trace over U(4) indices of the operator in (\ref{rho}),
\begin{equation}\label{chargedensity}
\hat\rho(\vec  x )=\sum_{\sigma=1}^4\sum_{\eta=1}^2
\frac{1}{2}\left[\Psi^{\sigma\dagger}_\eta( \vec x ),\Psi^\sigma_\eta(\vec x )\right]
\end{equation}
The interaction potential $V(\vec x-\vec y)$ is assumed to be instantaneous and translation invariant.
We shall also assume that it is a positive definite
kernel, that is, that its Fourier transform $\tilde V(\vec k)=\int d^2x e^{-i\vec k\cdot\vec x}V(\vec x)$
is non-negative, $\tilde V(\vec k)\geq 0$. This means that the potential is repulsive at all distance
scales.

Here, we have used the continuum, low energy theory of graphene which has explicit $U(4)$ symmetry.  Of course
this symmetry is not exact but is emergent at wavelengths much larger than the lattice scale.  In the action
which describes our model, the free field term
$$S_0=\int dt\int d^2x ~~i\bar\psi^\sigma\gamma^\mu D_\mu\psi^\sigma $$
has SO(2,1) relativistic invariance as well as U(4), whereas the interaction terms\footnote{The  nonlocal nature of the Coulomb interaction is due to the fact that the electromagnetic fields extend
out of the graphene plane into three space  dimensions where
 \begin{equation}\label{Coulomb}
V_{\rm Coulomb}(\vec x)= \frac{e^2}{4\pi\epsilon |\vec x|}=\frac{1}{4\pi\epsilon}\frac{1}{\sqrt{\vec
\nabla^2}}\delta(\vec x)
\end{equation}
The
Coulomb interaction is approximately instantaneous since
the speed of light in vacuum is about 300 times larger than the Fermi
velocity, which is the emergent  speed of light.
The Coulomb potential contains no dimensional parameters
and is therefore scale invariant. The quantum field theory defined by (\ref{fullhamiltonian}) with $V$ the
Coulomb potential (\ref{Coulomb})
is scale invariant at the classical level.}
$$S_{\rm Coulomb}=-\frac{e^2}{8\pi\epsilon}\int dt\int d^2x ~ \psi^{\sigma\dagger}\psi^\sigma \frac{1}{\sqrt{-\vec\nabla^2}}\psi^{\sigma'\dagger}\psi^{\sigma'}$$
is U(4) but is not relativistically invariant.  In spite of this, it is known that, in the absence
of background magnetic field, the field theory with action $S=S_0+S_{\rm Coulomb}$ is renormalizable
in the sense that ultraviolet divergences encountered in diagrammatic perturbation theory
can be canceled by adding counter-terms with the same form as those
operators already appearing
in the action.   The operators $i\bar\psi^\sigma \gamma^0D_t\psi^\sigma$ and $i\bar\psi^\sigma \vec\gamma
\cdot \vec D\psi^\sigma$ obtain different coefficients and the speed of light has logarithmic renormalization and
becomes a scale-dependent parameter where it runs to larger values at low
energies and momenta~\cite{gonzalez}-\cite{vafek}. The graphene fine structure constant $\alpha_{\rm graphene}= \frac{e^2}{4\pi\epsilon_0\hbar v_F}$, which controls the strength of the interaction, then runs to smaller values
 in the infrared, making the weakly coupled field theory stable under renormalization.
The Coulomb interaction in graphene is thought to be strong, posing a difficulty for the validity of this
perturbative approach, which would by valid if, for example, there were some screening mechanism which makes
the Coulomb interaction weaker than the naive coupling $\alpha_{\rm graphene}=\alpha\frac{c}{v_F}\approx \frac{300}{137}$.   It is fair to say that there is presently no quantitative picture of how such
a screening mechanism would work.  In this Paper, we shall also take the assumption that the interaction is weak
and we explore the consequences.  Though being perturbative, our techniques are not quite as elegant as the Feynman
diagrams which are used in Refs.~\cite{gonzalez}-\cite{vafek} since we must first deal with the vacuum degeneracy
which occurs at the leading order in perturbation theory.

 The interaction in (\ref{fullhamiltonian}) was modeled on this Coulomb interaction which, when only
components of the fermion operators which have wave-vectors very close to the $K$ and $K'$ degeneracy points are kept, have
the U(4) symmetric form with charge density given in (\ref{chargedensity}).
It is possible to keep track of the leading explicit
$U(4)$ symmetry breaking terms in the Hamiltonian, which arise from the lattice scale Coulomb interaction. They
appear as higher dimensional local operators that are
suppressed by powers of the lattice constant.  This was done by Alicea and Fischer \cite{alicea}. It is also possible
that there are other short ranged interactions that are compatible with the lattice symmetry and which do not have the form of
a charge density-charge density interaction that we are considering. An important example is the electron-phonon interaction
which has recently been discussed in Ref.~\cite{kharitonov}.
Generally, in the continuum limit, such interactions are also suppressed by
positive powers of the lattice spacing and are therefore small.  We will ignore all of
these corrections  while we study the problem of spontaneous $U(4)$ symmetry breaking in the
model (\ref{fullhamiltonian}). The symmetry breaking terms, as well as some others, for example the Zeeman interaction
which, due to the presence of the external magnetic field, breaks the spin symmetry can become important later,
in that they can determine the $U(4)$ orientation of the condensate once the symmetry is broken. This can, for example, favor
a truly ferromagnetic state with all spins aligned with the magnetic field over  a sublattice symmetry breaking
antiferromagnetic state.
 In spite of important differences in their physical properties, in
the $U(4)$ invariant model, these states would be related by $U(4)$ rotations and would
therefore appear to be indistinguishable.

In this Paper, we are going to find that, if the interaction strength
is sufficiently weak, and when the interaction is positive and couples
charge densities as in (\ref{fullhamiltonian}), the quantum Hall ferromagnet state, defined in Eq.~(\ref{qhf}) and denoted by $\left|{\rm qhf}\right>$, is the ground state of (\ref{fullhamiltonian}).  $\left|{\rm qhf}\right>$ is an eigenstate of the
single-particle Hamiltonian $H_0$ and has a particular distribution of electrons in the half-filled Dirac point Landau level, and with all Landau levels below the Dirac point filled and all above the Dirac point empty.
When interactions are taken into account,
$\left|{\rm qhf}\right>$ is not the exact ground state.  It is corrected in  perturbation theory by the usual formula,
\begin{equation}\label{corrected}
|{\rm gs}>~ =~ |{\rm qhf}>~-~\frac{\cal P}{H_0-E_0}~\frac{1}{2}\int d^2xd^2yV(\vec x-\vec y)~\rho(\vec x)\rho(\vec y)~|{\rm qhf}>
+\ldots
\end{equation}
which mixes it with higher Landau levels (the second term in (\ref{corrected}) is a superposition of single-particle
states, all of which have particles and holes occupying higher Landau levels).


In the following, we prove that the state (\ref{corrected}) is the ground state of (\ref{fullhamiltonian}) when
the strength of the interaction is sufficiently weak that perturbation theory is accurate. This proof includes the
contributions of all Landau levels.
This is an extension of previous results where this was demonstrated  in a lowest Landau level approximation \cite{qhf}\cite{qhf2}.  Indeed, for the scale invariant Coulomb interaction,
all Landau levels do contribute to the interaction energy at the same order of magnitude and Landau level mixing could
be a strong effect.


\section{The Quantum Hall Ferromagnet State}

In this section, we will construct the quantum Hall Ferromagnet state using the second quantized formalism.
First we need to study the single particle states in a magnetic field.
In the symmetric gauge, where $\vec A(\vec x) = -\frac{B}{2}\left( y,-x\right)$,
the single-particle Hamiltonian (\ref{hamiltonian}) can be written as
\begin{equation}
h=\left[ \begin{matrix} 0 &\sqrt{2B}\alpha^\dagger \cr \sqrt{2B}\alpha & 0 \cr \end{matrix}\right]
\end{equation}
where we have assumed that $B>0$ and we have introduced oscillator operators
\begin{eqnarray}\label{oscillatoralpha}
\alpha =\tfrac{1}{\sqrt{2B}}\left[ i\partial_x-\partial_y-\tfrac{B}{2}(y-ix)\right] ~~,~~
\alpha^\dagger = \tfrac{1}{\sqrt{2B}}\left[i\partial_x+\partial_y-\tfrac{B}{2}(y+ix)\right]\\
 \beta  =\tfrac{1}{\sqrt{2B}}\left[ i\partial_x+\partial_y+\tfrac{B}{2}(y+ix)\right] ~~,~~
 \beta^\dagger =\tfrac{1}{\sqrt{2B}}\left[ i\partial_x-\partial_y+\tfrac{B}{2}(y-ix) \right]
\end{eqnarray}
with non-vanishing commutators
$[\alpha,\alpha^\dagger]=1$ and $[\beta,\beta^\dagger]=1$.
Note that $\beta$ and $\beta^\dagger$ do not appear in the Hamiltonian. They
contain the guiding center coordinates.
The single-particle energies are obtained by solving the eigenvalue problem
\begin{equation}\label{eigenvalueproblem}
h\psi_{nm}(\vec x) = E\psi_{nm}(\vec x)
\end{equation}
The oscillator ground state is the Gaussian function
\begin{equation}
f_0(\vec x)=f_0(x,y)=\frac{B}{2\pi} e^{-\frac{B}{4}(x^2+y^2)}~:~\alpha f_0=0~~,~~\beta f_0=0
\end{equation}
We use this function to construct
an infinite tower of zero modes of (\ref{eigenvalueproblem}),
\begin{equation}\label{zeromodes}
\psi_{m}(\vec x)=\left[ \begin{matrix}u_m(\vec x)\cr 0\cr \end{matrix}\right]
=\left[ \begin{matrix} \frac{(\beta^\dagger)^m}{\sqrt{m!}}f_0(\vec x) \cr 0\cr \end{matrix}\right] ~~,~~
m=0,1,2,...
~~,~~E=0
\end{equation}
and positive energy (electron) and negative energy (hole) Landau level states
\begin{equation}\label{particleholewavefunctions}
\psi_{nm}^{(\pm)}(\vec x)=\frac{1}{\sqrt{2}}\left[ \begin{matrix} u_{nm}(\vec x)\  \cr
\pm u_{(n-1)m}(\vec x)\cr \end{matrix}\right]
 ~~
n=1,2,\ldots;~ m=0,1, ...~~,~~ E=\pm\sqrt{2Bn}
\end{equation}
where
\begin{equation}\label{particleholewavefunctions}
u_{nm} (\vec x)=  \frac{(\alpha^\dagger)^{n}}{\sqrt{n!}}
\frac{(\beta^\dagger)^m}{\sqrt{m!}}f_0  (\vec x)
\end{equation}

The second-quantized electron field is,
\begin{equation}\label{secondquantized}
\Psi^{ \sigma}(\vec x,t )=\sum_{m=0}^\infty\psi_{m}(\vec x)\vartheta_{ m}^{ \sigma} + \sum_{m=0}^\infty\sum_{n=1 }^\infty
\left( \psi_{nm}^{(+)}(\vec x) a_{nm}^{ \sigma}
+\psi_{nm}^{(-)}(\vec x) b_{nm}^{ \sigma\dagger}\right)
\end{equation}
The non-vanishing anti-commutators of zero mode, electron and hole creation and annihilation operators are
\begin{equation}
\left\{ \vartheta_{ m}^{ \sigma}, \vartheta_{ m'}^{ \sigma'\dagger}\right\} = \delta_{mm'}
\delta^{\sigma\sigma'}
,~
\left\{  a_{nm}^{ \sigma},  a_{n'm'}^{ \sigma'\dagger}\right\} = \delta_{mm'}\delta_{nn'}
\delta^{\sigma\sigma'}
,~
\left\{ b_{nm}^{ \sigma}, b_{n'm'}^{ \sigma'\dagger}\right\} = \delta_{mm'}\delta_{nn'}
\delta^{\sigma\sigma'}
\end{equation}
respectively.  With completeness sums for wave-functions, these guarantee that the second quantized
electron operator (\ref{secondquantized}) obeys the anticommutator (\ref{fieldanticommutator}).
The non-interacting part of the Hamiltonian is diagonal in the single-particle states.  Its second
quantized form is
\begin{equation}
H_0 =\sum_{n=1}^\infty\sum_{m=0}^\infty\sum_{\sigma=1}^4
\sqrt{2Bn}\left(a_{nm}^{\sigma\dagger} a^\sigma_{nm}
+b_{nm}^{\sigma\dagger} b^\sigma_{nm}\right) +E_0
\end{equation}
Note that the zero mode creation and annihilation operators do not appear in this expression.
$E_0$ denotes the ground state energy, which is equivalent to the sum over zero point energies of the electron and hole
states.
The total electric charge is given by
\begin{equation}
Q =\int dx \sum_{\sigma=1}^4\rho^{\sigma\sigma}(\vec x)=\sum_{n=1}^\infty\sum_{m=0}^\infty\sum_{\sigma=1}^4
\left(a_{nm}^{\sigma\dagger} a^\sigma_{nm}
-b_{nm}^{\sigma\dagger} b^\sigma_{nm}\right) +\sum_{m=0}^\infty\sum_{\sigma=1}^4
\left(\vartheta_{m}^{\sigma\dagger} \vartheta^\sigma_{n}-\frac{1}{2}\right)
\end{equation}
As usual, electrons and holes have positive and negative charges. The zero-modes on the other hand
carry fractional charge $\pm\frac{1}{2}$ per degree of freedom.
This is familiar from zero modes in other contexts, such as fractionally
charged solitons \cite{polyacetylene2}\cite{polyacetylene3}, and was discussed in the context of graphene in Ref.~\cite{Semenoff:1984dq}, but will not play an important role here, except to illustrate
that charge neutrality requires that half of the zero-mode states must be filled.

In perturbation theory, which we shall employ to study the full Hamiltonian $H_0+H_I$,
we begin by studying the degenerate neutral ground states of $H_0$.
A ground state of   $H_0$   has  all of the non-zero energy
particle and hole states empty, that is, a ground  state is annihilated by the particle and
hole annihilation operators
$ a_{nm}^{ \sigma} $ and $ b_{nm}^{ \sigma} $.
In order to be electrically neutral, it should also have half of the zero modes occupied.
 This leads to a large degeneracy of the ground state corresponding to all choices of half-filling of the zero modes.
Our task will be to study the leading order in degenerate perturbation theory to determine which of
 the half-filled states has the lowest energy once interactions are taken into account.

In the next Section, we shall prove that, amongst all possibilities,
a candidate for a ground state that it chooses
is  the   ``quantum Hall ferromagnet'' (qhf) state,  which is one of an infinite family of states, an
 example of which is
\begin{equation}\label{qhf}
\left|{\rm qhf}\right>
=\prod_{m }  \left( \vartheta^{1\dagger}_{m} \vartheta^{2\dagger}_{m}\right)\left|0\right>
\end{equation}
Here, $\left|0\right>$ is the state with no excited electrons or holes,
$ a_{nm}^{ \sigma}\left|0\right>=0$, $b_{nm}^{ \sigma}\left|0\right>=0$ and the zero energy Landau level empty,
$\vartheta_{m}^{ \sigma}\left|0\right>=0$ for $m=0,1,2,...$ and $n=1,2,...$.
To half-fill the zeroth Landau level, we must create two electrons for each label (m).
The creation operators in (\ref{qhf}) do just that.  Another way to specify this state, which avoids the infinite
product of creation operators, is
to have it satisfy
\begin{equation} a_{nm}^{ \sigma}\left|{\rm qhf}\right>=0~, ~b_{nm}^{ \sigma}\left|{\rm qhf}\right>=0~,~
\vartheta_{m}^{ \sigma=1,2\dagger}
\left|{\rm qhf}\right>=0~,~ \vartheta_{m}^{ \sigma=3,4}\left|{\rm qhf}\right>=0
\label{qhfdefn}\end{equation}
Note that we have chosen to create states with $U(4)$ labels
1,2.  In fact, we could have chosen any of a family of states which are related to this one by $U(4)$ transformations.
The most general such state is
\begin{equation}
\left|{\rm qhf}:~c~\right>
=\prod_{m }  \left(\sum_{\sigma<\sigma'} c_{\sigma\sigma'}\vartheta^{\sigma\dagger}_{m} \vartheta^{\sigma'\dagger}_{m}\right)\left|0\right>
\end{equation}
where, $\sum_{\sigma<\sigma'}|c_{\sigma\sigma'}|^2=1$.  Here, $c_{\sigma\sigma'}$ is independent of $m$.
For each $m$, there is an identical superposition of the six basis vectors
$\vartheta^{ \sigma\dagger}_{m}\vartheta^{ \sigma'\dagger}_{m}|0>$ (with the six
possible pairs of non-identical indices $(\sigma,\sigma')$), which
are a basis for the six-dimensional fundamental representation of the
U(4) algebra represented by the Young tableau which has a single column with two boxes.

If the $\left|{\rm qhf}\right>$ state is the ground state, it breaks the $U(4)$ symmetry to a residual $U(2)\times U(2)$,
those transformations that are in $U(4)$ and which still leave the state in (\ref{qhf}) unchanged.
The zero-mode parts of the generators of the $U(2)\times U(2) $ residual symmetry
subalgebra of $U(4)$
which annihilate $\left|{\rm qhf}\right>$, and therefore generate unbroken symmetries  are
\begin{align}
J^1=\frac{1}{2} \left(\vartheta_m^{1\dagger}\vartheta_m^1-\vartheta_m^{2\dagger}\vartheta_m^2\right)  ,
J^2=\frac{1}{2} \left(\vartheta_m^{1\dagger}\vartheta_m^2+\vartheta_m^{2\dagger}\vartheta_m^1\right)  ,
J^3=\frac{1}{2} \left(-i\vartheta_m^{1\dagger}\vartheta_m^2+i\vartheta_m^{2\dagger}\vartheta_m^1\right)
\nonumber\\
\tilde J^1=\frac{1}{2} \left(\vartheta_m^{3\dagger}\vartheta_m^3-\vartheta_m^{4\dagger}\vartheta_m^4\right)  ,
\tilde J^2=\frac{1}{2} \left(\vartheta_m^{3\dagger}\vartheta_m^4+\vartheta_m^{4\dagger}\vartheta_m^3\right)  ,
\tilde J^3=\frac{1}{2} \left(-i\vartheta_m^{3\dagger}\vartheta_m^4+i\vartheta_m^{4\dagger}\vartheta_m^3\right)
\nonumber\\
\tilde Q=  \vartheta_m^{1 }\vartheta_m^{1\dagger}+\vartheta_m^{2}\vartheta_m^{2\dagger}  +
\vartheta_m^{3\dagger}\vartheta_m^3+\vartheta_m^{4\dagger}\vartheta_m^4  ,~
Q=  -\vartheta_m^{1 }\vartheta_m^{1\dagger}-\vartheta_m^{2}\vartheta_m^{2\dagger}   +
\vartheta_m^{3\dagger}\vartheta_m^3+\vartheta_m^{4\dagger}\vartheta_m^4
\label{residual}\end{align}
where $m=0,1,2,...$ is summed in each term and
we have omitted similar terms with particle and hole creation and annihilation operators in each
generator.

\section{Degenerate Perturbation Theory}

We have studied some of the kinematical properties of the $\left|{\rm qhf}\right>$ state.  Now,
we must establish that, if the interactions are weak enough, $\left|{\rm qhf}\right>$  is
indeed a ground state.  We shall do this by demonstrating that it saturates a lower bound
on the expectation value of the interaction Hamiltonian.  This shows that it is one of a family
of degenerate ground states which are related to each other by $U(4)$ transformations.  We note here
this is already well known in an approximation where the mixing of Landau levels is ignored~\cite{qhf1}.
The novel aspect of our demonstration is that it takes the mixing of higher Landau levels into account.
The fact that the contribution
of higher Landau levels cancels is essentially a consequence of the special symmetries of the charge neutral
state of graphene and will not apply to the generic quantum Hall ferromagnet states of bilayer semiconductors.

The Hamiltonian (\ref{fullhamiltonian}), written in terms of creation and annihilation
operators, is\footnote{The zero-point energy of the non-interacting Dirac electrons
can be defined by Zeta-function regularization,
\begin{equation}\label{energy}
E_0=-\frac{1}{2\pi}\sqrt{2}\zeta\left(-  \tfrac{1}{2} \right)|B|^{\frac{3}{2}}V
\end{equation}
Here, $V$ is the volume
of the two dimensional space.
We have defined the infinite sum over energies of Landau levels
as the analytic continuation of Riemann's
zeta function,
\begin{equation}\zeta(-\tfrac{1}{2})=\lim_{s\to -\tfrac{1}{2}}\sum_{n=1}^\infty n^{-s}=-.2078862250...
\end{equation}
Note that the zeta function is finite at $s=-\tfrac{1}{2}$.
The $|B|^{\frac{3}{2}}$ behavior of the energy density is due to the
scale invariance of massless fermions. Once
the negative sign of the zeta function is taken into account,
the ground state energy (\ref{energy}) is positive.}
\begin{eqnarray}
H=E_0(B)+\sum_{nm\sigma } \sqrt{2B n }\left(a^{\sigma\dagger}_{nm} a_{nm}^{\sigma} +
b^{\sigma\dagger}_{nm}b_{nm}^{\sigma}\right) +
\frac{1}{2}\int d^2x d^2y \hat\rho(\vec x)V(\vec x - \vec y) \hat\rho(\vec y)
\end{eqnarray}
We will find it convenient to split the charge density operator into three parts,
\begin{equation}\label{totalrho}
\hat\rho(\vec x) = \sum_{ \sigma=1}^4\frac{1}{2}\left[ \Psi^{
\sigma\dagger}(\vec x),\Psi^{ \sigma}(\vec x)\right] ~\equiv~ \hat\rho_1(\vec x)+\hat\rho_2(\vec x)+\hat\rho_3(\vec x)
\end{equation}
where $\hat\rho_1(\vec x)$ is quadratic in $\vartheta_m$ and $\vartheta_m^\dagger$,
\begin{align}
\hat\rho_1(\vec x)=\sum_{mm'=0}^\infty \psi_m^\dagger(\vec x)\psi_{m'}(\vec x)\sum_{\sigma\sigma'=1}^4\left[\theta^{\sigma\dagger}_m\theta^\sigma_{m'}
-\frac{1}{2}\delta_{mm'}\right]
\label{rhoone}
\end{align}
$\hat\rho_2(\vec x)$ is linear in
$\vartheta_m$ and $\vartheta_m^\dagger$,
\begin{align}
\hat\rho_2(\vec x)=\sum_{n=1}^\infty\sum_{mm'=0}^\infty\sum_{\sigma\sigma'=1}^4 &\left(\left[a^{\sigma\dagger}_{nm}\psi_{nm}^{(+)\dagger}(\vec x)
+
b_{mn}^\sigma\psi_{nm}^{(-)\dagger}(\vec x)\right]\psi_{m'}^{(+)}(\vec x)\theta^{\sigma }_{m'}
+  \right.
\nonumber \\
&\left. +
\theta^{\sigma\dagger}_{m'}\psi^\dagger_{ m'} (\vec x)
\left[\psi_{nm}^{(+) }(\vec x) a^{\sigma }_{nm}
+
\psi_{nm}^{(-) }(\vec x)b_{mn}^{\sigma\dagger}\right]  \right)
\label{rhotwo}
\end{align}
and $\hat\rho_3(\vec x)$ is
independent of $\vartheta_m$ and $\vartheta_m^\dagger$,
\begin{align}
\hat\rho_3(\vec x)=\sum_{nn'=1}^\infty\sum_{mm'=0}^\infty\sum_{\sigma\sigma'=1}^4 :\left[a^{\sigma\dagger}_{nm}\psi_{nm}^{(+)\dagger}(\vec x)
+
b_{mn}^\sigma\psi_{nm}^{(-)\dagger}(\vec x)\right]
\left[\psi_{n'm'}^{(+) }(\vec x) a^{\sigma }_{n'm'}
+
\psi_{n'm'}^{(-) }(\vec x)b_{m'n'}^{\sigma\dagger}\right]:
\label{rhothree}
\end{align}
where $:...:$ denote normal ordering.

Perturbation theory is complicated by the fact that $H_0$ has a degenerate ground state. $E_0$ is
its smallest eigenvalue and
the eigenvalue equation $
H_0 \left|c\right> = E_0 \left|c\right>
$
is satisfied by any state   $\left|c\right>$ which is annihilated by
both particle and hole annihilation operators,   $a_{nm}^\sigma\left|c\right>=0$ and $b_{mn}^\sigma\left|c\right>=0$ for all  $m,n,\sigma$.
These states can contain any configuration of occupations of the zero modes. For the following
argument, we need not
restrict ourselves to charge neutral states only.

As  is usual in degenerate perturbation theory, the degeneracy of these states is resolved
by finding the eigenvalues of the matrix
$
 \left<c\right|H_I\left|c'\right>
$
in the space of degenerate ground states of $H_0$.
Eigenvalues of the matrix $\left<c\right|H_I\left|c'\right>$
are the energies of the lowest eigenstates of $H=H_0+H_I$ to the leading order in degenerate perturbation
theory.  Of course, finding the complete spectrum of eigenvalues is a difficult problem which
we shall not be able to solve here.  We will, however, be able to identify the ground states.
By the standard variational argument, the eigenvectors with the lowest eigenvalues can be found
by searching for a state $\left|c\right>$  for which
$\left<c\right| H_I \left|c\right>$ is a minimum.  Such a state is guaranteed to be a ground state, or one
of a set of degenerate ground states.

If  $\left|c\right>$ is any of the possible ground states of $H_0$, it is easy to see that
\begin{align}
 \left<c\right|H_I\left|c\right> &= \frac{1}{2}\int d^2x  d^2y V(\vec x-\vec y)
\left<c\right|\hat\rho(\vec x)\hat\rho(\vec y)\left|c\right>
\nonumber \\ & =\frac{1}{2}\int d^2x  d^2y V(\vec x-\vec y)
\sum_{i=1}^3\left<c\right|\hat\rho_i(\vec x)\hat\rho_i(\vec y)\left|c\right>
\end{align}
where $\hat\rho_i(\vec x)$ are the terms in (\ref{totalrho}), given in Eqs.~(\ref{rhoone}), (\ref{rhotwo}) and
(\ref{rhothree}).
We can treat the terms with $\left<c\right|\hat\rho_i(\vec x)\hat\rho_i(\vec y)\left|c\right>$ separately for each $i$.

Let us begin with $\left<c\right|\hat\rho_1(\vec x)\hat\rho_1(\vec y)\left|c\right>$, where
$\hat\rho_1(\vec x)$ is given in Eq.~(\ref{rhoone}).
First of all, it is straightforward to see that $\left|{\rm
qhf}\right>$ is an exact eigenstate of the density $\hat\rho_1(\vec x)$
with zero eigenvalue,
\begin{equation}
\hat\rho_1(\vec y)\left|{\rm qhf}\right>=0
\end{equation}
so that $\frac{1}{2}\int d^2x  d^2y V(\vec x-\vec y)\left<{\rm qhf}\right|\hat\rho_1(\vec x)\hat\rho_1(\vec y)\left|{\rm qhf}\right>=0$ is indeed the minimum if $V(\vec x -\vec y)$ is a positive kernel\footnote{
If $\hat\rho_1(k)=\frac{d^2x}{2\pi}e^{-i\vec k\cdot\vec x}\hat\rho_1(\vec x)$
is the Fourier transform of $\hat\rho_1(x)$,
and $V(\vec x)=\int\frac{d^2k}{(2\pi)^2}e^{i\vec k\cdot
\vec x}V(k)$,
 \begin{align}
\frac{1}{2}\int d^2x  d^2y V(\vec x-\vec y)\left<c\right|\hat\rho_1(\vec x)\hat\rho_1(\vec y)\left|c\right>
=\left<c\right| ~\int d^2k \hat\rho_1(-\vec k)V(\vec k)\hat\rho_1(\vec k) ~\left|c\right>=\nonumber \\
=\sum_{\Theta} \int d^2k \left| \left<\Theta\right|\hat\rho_1(\vec k)\left|c\right>\right|^2 V(\vec k)
\geq 0
\end{align}
where $\{\left|\Theta\right>\}$ are a complete set of intermediate states.
This inequality is saturated by $\left|c\right>=\left|{\rm qhf}\right>$, since, as we shall show,
$\hat\rho_1(\vec x)\left|{\rm qhf}\right>=0$.}.

To see that $\hat\rho_1(\vec x)\left|{\rm qhf}\right>=0$, we observe that,
when we operate $\hat\rho_1(\vec x)$ on
the $\left|{\rm qhf}\right>$ state defined in (\ref{qhf}) we obtain
\begin{equation}\label{rhooneproof}
\hat\rho_1(\vec x) \left|{\rm qhf}\right>
=\sum_{mm'}\psi_m^{\dagger}(\vec x)
\psi_{m'}(\vec x)\sum_{\sigma=1}^4\left(
\theta^{\sigma\dagger}_{m}\vartheta^{\sigma}_{m'}-\frac{1}{2}\delta_{mm'} \right)
\prod_{\tilde m'' }  \left( \vartheta^{1\dagger}_{\tilde m''} \vartheta^{2\dagger}_{\tilde m''}\right)\left|0\right>
\end{equation}
To evaluate this expression, it is easiest to separate the summation over $m$ and $m'$  into two parts,
the first consisting of
the terms where $m=m'$ and the second consisting of the terms where $m\neq m'$.

Consider the terms with $m=m'$. The state $|{\rm qhf}>$ is
an eigenstate of each term $\left(
\theta^{\sigma\dagger}_{m}\vartheta^{\sigma}_{m}-\frac{1}{2} \right)$ with eigenvalue $\frac{1}{2}$ for $\sigma=1,2$
and eigenvalue $-\frac{1}{2}$ for $\sigma=3,4$. When summed over $\sigma$, these contributions cancel, for each value of $m$.
Therefore, the terms with $m=m'$ have vanishing contribution in (\ref{rhooneproof}).

Now, consider the terms where $m\neq m'$ which are of the form $\theta^{\sigma\dagger}_{m}\vartheta^{\sigma}_{m'}$.
They annihilate a zero mode in state $m'$ and create one in a different state $m$, but without
changing the $\sigma$-label.  However, because of Fermi statistics,
this is only possible when the $m'$-state is occupied and the $m$-state with the same $\sigma$-label
is unoccupied.  The $\left|{\rm qhf}\right>$ state is specifically constructed so that all  states with a given $\sigma$-label
are either completely filled or completely empty,
so this process is not allowed in the $\left|{\rm qhf}\right>$ state and the terms with $m\neq m'$ must therefore also vanish in (\ref{rhooneproof}). Thus we conclude that $\hat\rho_1(\vec x)\left|{\rm qhf}\right>=0$.

If we truncated the interaction to the lowest (charge neutral) Landau level, terms containing $\hat\rho_2(\vec x)$ and $\hat\rho_3(\vec x)$ would be dropped from the interaction Hamiltonian and we could stop here.
We would conclude that $\left|{\rm qhf}\right>$ is
an eigenstate of the interaction Hamiltonian $H_I$   with zero eigenvalue, saturates a lower bound on
the energy and
it is therefore a  ground state.  We will now include the mixing of
the non-zero Landau levels.  This mixing is due to the presence of $\hat\rho_2(\vec x)$ and $\hat\rho_3(\vec x)$ in
the interaction Hamiltonian.  We will show that, even when this mixing is included, $\left|{\rm qhf}\right>$ remains the lowest energy state.

To this end, we have already shown that $\frac{1}{2}\int d^2x  d^2y V(\vec x-\vec y)
 \left<{\rm qhf}\right|\hat\rho_1(\vec x)\hat\rho_1(\vec y)\left|{\rm qhf}\right>=0 $ and saturates the lower bound on
the positive semi-definite expectation value $\frac{1}{2}\int d^2x  d^2y V(\vec x-\vec y)
 \left<c\right|\hat\rho_1(\vec x)\hat\rho_1(\vec y)\left|c\right>  $ where $\left|c\right>$ is any of the array of possible ground states
of $H_0$, that is, any state which has no particles or holes in non-zero energy Landau levels and has any fractional filling of
the charge neutral point, zero energy Landau level.  We will now show that
the other two terms in the expectation value of the interaction Hamiltonian, $\frac{1}{2}\int d^2x  V(\vec x-\vec y)
 \left<c\right|\hat\rho_2(\vec x)\hat\rho_2(\vec y)\left|c\right> $ and
$\frac{1}{2}\int d^2x  d^2y V(\vec x-\vec y)
 \left<c\right|\hat\rho_3(\vec x)\hat\rho_3(\vec y)\left|c\right> $
are independent of which state $\left|c\right> $ is used to take the matrix element, as long as it has no excited particles or holes.  This will be a result
of CP symmetry.

Consider the interaction with density $\hat\rho_2(\vec x)$ which, when we operate on   $\left|c\right>$, produces
states with a single particle or a single hole in a higher Landau level,
\begin{equation}
 \hat\rho_2(x)\left|c\right>=\sum_{mm'n}
\psi^{(+)\dagger}_{nm}(x)\psi_{m'}(x)
a_{nm}^{\sigma\dagger}\vartheta_{m'}^\sigma \left|c\right>
+ \sum_{mm'n}
\psi^\dagger_{m'}(x)\psi^{(-)}_{nm}(x)
\vartheta_{m'}^{\sigma\dagger} b_{nm}^{\sigma\dagger} \left|c\right>
\end{equation}
 and
\begin{align}
&\int d^2xd^2y~V(x-y) \left<c\right|\hat\rho_2(x)\hat\rho_2(y)\left|c\right>= \nonumber \\ &=
\int d^2xd^2y~V(x-y)\sum_{nmm'}\psi^{\dagger}_{m'}(x) \psi^{(+)}_{nm}(x)  \psi^{(+)\dagger}_{nm}(y)\psi_{m'}(y)
\left<c\right|\sum_\sigma\vartheta_{m'}^{\sigma\dagger} \vartheta_{m'}^\sigma \left|c\right>
\nonumber \\ &+
\int d^2xd^2y~V(x-y)\sum_{nmm'}\psi^{\dagger}_{m'}(x) \psi^{(-)}_{nm}(x)  \psi^{(-)\dagger}_{nm}(y)\psi_{m'}(y)
\left<c\right|\sum_\sigma \vartheta_{m'}^\sigma \vartheta_{m'}^{\sigma\dagger}\left|c\right>
\label{1234}\end{align}
Now, we note that, because $\psi^{\dagger}_{m'}(x) \psi^{(-)}_{nm}(x) = \psi^{\dagger}_{m'}(x) \psi^{(+)}_{nm}(x)$,  the wave-function contributions from intermediate
particles and holes are identical in the above equation. This allows us to combine terms so that
the zero mode creation and annihilation operators in (\ref{1234}) combine as an anti-commutator.   The result is
\begin{align}
&\int d^2xd^2y~V(x-y) \left<c\right|\hat\rho_2(x)\hat\rho_2(y)\left|c\right>=\nonumber \\ &~~~~~=
4\int d^2xd^2y~V(x-y)\sum_{nmm'}\psi^{\dagger}_{m'}(x) \psi^{(+)}_{nm}(x)  \psi^{(+)\dagger}_{nm}(y)\psi_{m'}(y)
\end{align}
The right-hand-side of this equation is independent of the state
 $\left|c\right>$ which is used on the left-hand-side.  This means that this
contribution to the energy is the same for any $\left|c\right>$.

Now, consider the third contribution from $\hat\rho_3(x)$ which, when operating on  $\left|c\right>$, creates a particle-hole
pair in higher Landau levels,
\begin{equation}
 \hat\rho_3(x)\left|c\right>= \sum_{mm'nn'}
\psi^{(+)\dagger}_{nm}(x)\psi^{(-)}_{n'm'}(x)
a_{nm}^{\sigma\dagger}b_{n'm'}^{\sigma\dagger} \left|c\right>
\end{equation}
The contribution to the energy is proportional to
\begin{align}
&\int d^2xd^2y~V(x-y) \left<c\right|\hat\rho_3(x)\hat\rho_3(y)\left|c\right>= \nonumber \\ &=~~~~~
4\int d^2xd^2y~V(x-y)\sum_{nmn'm'}\psi^{(-)\dagger}_{n'm'}(x) \psi^{(+)}_{nm}(x)  \psi^{(+)\dagger}_{nm}(y)\psi^{(-)}_{n'm'}(y)
\end{align}
This result is also independent of the population of zero modes in $\left|c\right>$.

Finally, we have demonstrated that
\begin{align}
&\left<c\right|H_I\left|c\right>=\frac{1}{2}\int d^2xd^2y~V(x-y)\sum_i \left<c\right|\hat\rho_i(x)\hat\rho_i(y)\left|c\right>=\nonumber \\ &=~~~~~
\frac{1}{2}\int d^2xd^2y~V(x-y)  \left<c\right|\hat\rho_1(x)\hat\rho_1(y)\left|c\right>
+\{{\rm ~independent~of~\left|c\right>}\}
\end{align}
Further, we have shown that the first term on the right-hand-side is minimal when $\left|c\right>=\left|{\rm qhf}\right>$. This establishes that $\left|{\rm qhf}\right>$ is indeed the state with lowest energy in degenerate perturbation theory.

Finally, we remark that $\left|{\rm qhf}\right>$ is an eigenstate of the Hamiltonian only at the zeroth
order of perturbation theory.  Once it has been identified as the state with the smallest expectation value
of the energy, we use it to construct the ground state that is corrected to first order in perturbation theory,
\begin{equation}
|{\rm gs}>~ =~ |{\rm qhf}>~-~\frac{\cal P}{H_0-E_0}~\frac{1}{2}\int d^2xd^2yV(\vec x-\vec y)~\rho(\vec x)\rho(\vec y)~|{\rm qhf}>
+\ldots\nonumber
\end{equation}
which we quoted in (\ref{corrected}),
where ${\cal P}$ is the projection onto states orthogonal to $|{\rm qhf}>$ and the dots indicate contributions of quadratic
and greater order in the interaction strength. The interaction Hamiltonian creates states with  electrons and holes
in higher Landau levels.

\section{Condensate}

The expectation value of the U(4) charge density in the perturbatively
corrected quantum Hall ferromagnet state (\ref{corrected}) is
\begin{align}\rho^{\sigma\sigma'}=&
<{\rm qhf}|\frac{1}{2}\left[ \Psi^{\sigma'\dagger}(\vec x ),  \Psi^{\sigma}(\vec x )\right]
 |{\rm qhf}> \nonumber \\ &  - \int d^2yd^2zV(\vec y-\vec z)
~{\rm Re}~<{\rm qhf}| \Psi^{\sigma'\dagger}(\vec x )  \Psi^{\sigma}(\vec x )
~\frac{\cal P}{H_0-E_0}\rho(\vec y)\rho(\vec z)|{\rm qhf}>+\ldots
\label{rhofinal}\end{align} and the chiral condensate is
\begin{align}\Sigma^{\sigma\sigma'}=&
<{\rm qhf}|\bar\Psi^{\sigma' }(\vec x )  \Psi^{\sigma}(\vec x )
 |{\rm qhf}> \nonumber \\ &  - \int d^2yd^2zV(\vec y-\vec z)
~{\rm Re}~<{\rm qhf}|\bar\Psi^{\sigma' }(\vec x )  \Psi^{\sigma}(\vec x )
~\frac{\cal P}{H_0-E_0}\rho(\vec y)\rho(\vec z)|{\rm qhf}>+\ldots
\label{sigmafinal}\end{align}
The leading terms on the right-hand-sides of
(\ref{rhofinal}) and (\ref{sigmafinal}) are governed strictly by the zero modes and
can easily be computed,
$$
\rho^{\sigma\sigma'}=\frac{1}{2}
{\rm diag}(1,1,-1,-1)\sum_{m=0}^\infty \psi_m^\dagger(x)\psi_m(x) =
{\rm diag}(1,1,-1,-1)\frac{|B|}{4\pi}
$$
$$
\Sigma^{\sigma\sigma'}=\frac{1}{2}{\rm diag}(1,1,-1,-1) \sum_{m=0}^\infty \psi_m^\dagger(x)\psi_m(x)+\ldots=
{\rm diag}(1,1,-1,-1)\frac{B}{4\pi} +\ldots
$$
which are the results quoted in (\ref{leadingorderrho}) and (\ref{leadingordersigma}) at the leading order.
To find this leading term, we have used the completeness sum quoted in Eq.~(\ref{uu}) below.

The second terms on the right-hand-side of  (\ref{rhofinal}) and (\ref{sigmafinal}) arise from interactions.
The perturbative corrections to the condensates in (\ref{rhofinal}) and (\ref{sigmafinal}) are quite different.
We can argue that the linear order correction to the U(4) charge density,
the second term on the left-hand-side of Eq.~(\ref{rhofinal}), vanishes. This is due to the
fact that,  because
of the projection operator, the only intermediate states that contribute are those with a particle-hole pair where both
are in non-zero Landau levels, or a particle or hole in a non-zero Landau level and a displaced zero mode. In both cases
the wave-functions which contribute are orthogonal -- they appear as $\psi^{(-)\dagger}_{nm}(x)\psi^{(+)}_{rs}(x)$
or $\psi^{(\pm)\dagger}_{nm}(x)\psi_{s}(x)$,  and their conjugates, where $mn$ and $rs$ are the Landau
level quantum numbers of the hole and the particle, respectively.  Because of orthogonality, the integral of this density over
space must vanish.  That is, the shift of the total U(4) charge, as opposed to the charge density, is zero.
However, because of translation invariance,
the matrix element must be an $\vec x$-independent constant.
It is thus a constant whose integral over space must vanish, it must
therefore be zero.
  Thus it applies to all terms at this first order perturbative computation of the U(4) charge density -- the entire correction to linear order in the interaction must vanish.

This argument can be extended to higher orders.  If the U(4) charge density is uniform, it can be deduced from knowledge of
the total U(4) charge which can be gotten from taking expectation values of the four operators
\begin{equation}
Q^\sigma =\sum_{n=1}^\infty\sum_{m=0}^\infty
\left(a_{nm}^{\sigma\dagger} a^\sigma_{nm}
-b_{nm}^{\sigma\dagger} b^\sigma_{nm}\right) +\sum_{m=0}^\infty
\left(\vartheta_{m}^{\sigma\dagger} \vartheta^\sigma_{n}-\frac{1}{2}\right)
\end{equation}
for each value of $\sigma$.  These operators commute with the free and interaction Hamiltonians (they are generators of
the U(4) symmetry) and they therefore
 commute with all of the operators which occur in perturbative correction of the $\left|{\rm qhf}\right>$ state.
This suggests that all perturbative corrections vanish and the   U(4) charge density which is computed in the leading order
is exact to all orders in perturbation theory.

The chiral condensate (\ref{sigmafinal}), on the other hand, obtains contributions from the linear order.
This contribution is from terms with  particle-hole intermediate
states where the analogous wave-functions appear as  $\psi^{(-)\dagger}_{mn}(x)\gamma^0\psi^{(+)}_{rs}(x)=\psi^{(-)\dagger}_{mn}(x)\psi^{(-)}_{rs}(x)$.  Integration of
this quantity over $\vec x$ gives $\delta_{mr}\delta_{ns}$, rather then zero.  This allows a contribution to the chiral
condensate.   It should, however by $\vec x$-independent and we can identify parts
which would be identical after we perform an integral.  In fact, an argument similar to the one that we used for
the U(4) charge operator suggests that we can replace the expectation value of the mass operator by the
expectation value of $$\frac{1}{v}\int d^2x~\frac{1}{2}\left[\bar\Psi^{\sigma'}(x),\Psi^\sigma(x)\right] =
\frac{1}{v}\sum_m\left(\theta^{\sigma'\dagger}_m\theta^\sigma_m-\frac{1}{2}\right)+\frac{1}{v}\sum_{mn}\left[ a_{mn}^{\sigma'\dagger} b_{mn}^{\sigma\dagger} +
b_{mn}^{\sigma'}a_{mn}^\sigma\right]$$
where $v$ is the volume of space.

To proceed, we need the sum rule (which we derive in the Appendix)
\begin{align}
\sum_{m =0}^\infty   u_{nm}(x)u_{nm}^*(y)
=
\tfrac{|B|}{2\pi}e^{-\tfrac{|B|}{4}(x-y)^2-i\tfrac{B}{2}x\times y}
L_n \left[\tfrac{|B|(x-y)^2}{2}\right]   \label{uu}
\end{align}
where, the summation is over guiding center labels and $L_n(x)$ are Laguerre polynomials, normalized so that the first few are
\begin{align}
L_0&=1
&L_1&=1-x  \nonumber \\
L_2&=1-2x+\tfrac{1}{2}x^2  &
L_3&=1 -3x+\frac{3}{2}x^2-\tfrac{1}{6}x^3
\nonumber
\end{align}
and
$$
L_n(x)=\frac{1}{n!}e^{x}\frac{d^n}{dx^n}(x^ne^{-x})
$$
With this sum rule, we find
\begin{align}
\Sigma =\frac{B}{4\pi}\left[1 +\frac{1}{4}\int_0^\infty dr \sqrt{\frac{2}{|B|}}V(\sqrt{2r/|B|})e^{-r}\sum_{p=1}^\infty\frac{L_p(r)}{\sqrt{p}} + {\cal O}(V^2)\right]
\end{align}
where, as indicated, corrections are at least of order two in the interaction strength $V$.

 For the Coulomb interaction
\begin{align}
\Sigma =\frac{B}{4\pi}\left[1 +\frac{e^2}{16\pi\epsilon} \sum_{p=1}^\infty\frac{(2p)!}{2^{2p}(p!)^2\sqrt{p}} + {\cal O}(V^2)\right]
\end{align}
For large $p$, the summand $\sim 1/(\sqrt{\pi} p)$ and the sum is logarithmically divergent.

\section{Conclusions}

In this Paper, we have shown that the weak coupling ground state (\ref{corrected})
of the continuum model of graphene (\ref{fullhamiltonian}) in a magnetic field
is the quantum Hall ferromagnet corrected by higher order terms. One of our results is the
systematic inclusion of  Landau level mixing.  We have shown that, because of the residual
particle-hole symmetry of the charge neutral state, Landau level mixing does not shift the
relative energies of the possible fractional fillings of the charge zero point Landau level.
The quantum Hall ferromagnet therefore remains the lowest energy state even after Landau level
mixing is taken onto account.

In order to understand the electron spectrum after interactions are turned on,
we can fit the expectation values of the bilinear fermion operators that we have found
to the same expectation values in a theory with free fermions.   There, one would expect that, as well
as the terms already present in $H_0$, 
the Hamiltonian which describes the fermions would acquire a term with a U(2)$\times$U(2) invariant
chemical potential $\int d^2x \Psi^{\sigma\dagger}(x)\mu_0^{\sigma\sigma'}(x-y)\Psi^{\sigma'}(y)$
where
\begin{align}\label{inducedmass}
\mu^{\sigma\sigma'}_0=\mu_0~{\rm diag}(1,1,-1,-1)
\end{align}
For reasons which we shall explain below, this chemical potential is not needed and it can be set to
zero.

In addition to a chemical potential, there could be a (non-local) U(2)$\times$U(2) invariant mass operator
$\int d^2x \bar\Psi^{\sigma}(x)M_0^{\sigma\sigma'}(x-y)\Psi^{\sigma'}(y)$ where
\begin{align}\label{inducedmass}
M_0^{\sigma\sigma'}=m_0~{\rm diag}(1,1,-1,-1)
\end{align}
and
\begin{align}
m_0^2(x-y)=\frac{2B\alpha\alpha^\dagger f^2(\alpha\alpha^\dagger)}
{1-f^2(\alpha\alpha^\dagger)}\delta(\vec x-\vec y)
\label{mass}
\end{align}
The differential operators $\alpha$ and $\alpha^\dagger$ are defined in Eq.~(\ref{oscillatoralpha}).
Note that they contain only covariant derivatives, so this mass operator is formally gauge covariant.
The spectrum of $\alpha\alpha^\dagger$ is the positive integers, $n=1,2,...$ and
\begin{align}\label{f(x)}
f(x)=\frac{e^2}{32\pi\epsilon} \frac{\Gamma[2x+1]}{2^{2x}(\Gamma[x+1])^2\sqrt{x}}
\end{align}
Computing the charge and chiral condensates in the charge neutral state
with the free fermion Hamiltonian with $\mu_0=0$ and $m_0$ given by (\ref{mass}) will
result in the condensates $\rho^{\sigma\sigma'}$ and $\Sigma^{\sigma\sigma'}$ that we
have found in the theory with interactions turned on.  Note that the induced mass is of linear and higher
orders  in the Coulomb coupling
$\frac{e^2}{4\pi\epsilon}$, making it small in the weak coupling limit where the analysis is valid.

It is interesting that, what began as a U(4) ferromagnetic state with a U(4) ferromagnetic charge density
is nominally a state with dynamical mass generation only.  The U(4) ferromagnetic charge density in this state
with dynamical mass generation simply results from the asymmetry of the single fermion spectrum with a mass
and in a magnetic field -- when a mass is present, the states which were zero modes of the massless Dirac Hamiltonian
and which occurred at the apex of the
Dirac cones are shifted and now occur at one of the mass thresholds, either $E=+m_0$ or $E=-m_0$, depending
on the sign of $B$.  There is the further fact that has been known for a long time~\cite{Niemi:1983rq}\cite{polyacetylene3},
that the spectra are otherwise symmetric -- all other states come in pairs of electron and hole states.  Thus we see that
the entire ``spectral asymmetry'' of the Dirac Hamiltonian comes from precisely the same states which were zero modes.
Furthermore, the charge of the single particle state is given by the spectral asymmetry~\cite{polyacetylene3}.  Therefore, the entire U(4) charge is due to these threshold modes.  Further, because
the interactions do not upset the symmetry of the remaining modes, this charge seems to survive perturbative
corrections.

In our detailed computations, we have used a shortcut where we deduce the charge density and the expectation value
of the chiral condensate by first considering the expectation values of the relevant densities integrated over
the volume of space.  The integral results in a much simpler operator and the expectation values are given by simple
formulae.   We then deduce the densities by dividing by the volume factor.  This is justified by the fact that the states
that we consider are invariant under translations.  The expectation value of charge densities should therefore be position
independent constants.  The total charge would be the integrals of these constants, which must then be equal to a volume
factor times the constant.  Particularly for the charge, where the integrals of corrections must vanish, we deduce that the
constant density should vanish.   One must caution the reader that there are cases where this proceedure does not work.  For
example, the summation over Landau level integers might not commute with doing a volume integral.  The non-normal ordered
chiral charge is one example.  The expectation value of the operator $\bar\Psi^1(x)\Psi^1(x)$ in the Hall ferromagnet state
would be
$$
\left<{\rm qhf}\right| \bar\Psi^1(x)\Psi^1(x) \left| {\rm qhf}\right> =\sum_{m=0}^\infty\psi_m^\dagger(x)\psi_m(x) + \sum_{n=1}^\infty\sum_{m=0}^\infty \psi^{(-)\dagger}_{nm}(x)\gamma^0\psi^{(-)}_{nm}(x)
$$
$$
= \sum_{m=0}^\infty\psi_m^\dagger(x)\psi_m(x) + \sum_{n=1}^\infty\sum_{m=0}^\infty
 \psi^{(-)\dagger}_{nm}(x)\psi^{(+)}_{nm}(x)
$$
If we integrate over $x$, due to the orthogonality of positive and negative energy states, the integral of the
second summation on the right-hand-side must vanish and we
would conclude that the total chiral charge density is given by the first term.
On the other hand, of we consider the density more explicitly,
\begin{align}
\left<{\rm qhf}\right| \bar\Psi^1(x)\Psi^1(x) \left| {\rm qhf}\right> &=\sum_{m=0}^\infty\psi_m^\dagger(x)\psi_m(x)
\nonumber \\
&+ \frac{1}{2}\sum_{n=1}^\infty\sum_{m=0}^\infty\left[  u_{nm}^{\dagger}(x)u_{nm}(x) - u_{(n-1)m}^\dagger(x)
u_{(n-1)m}(x)\right]
\nonumber\end{align}
and then we cancel terms which have the same lower indices in the second summation, we see that a term
with zero mode indices does not cancel and we obtain
$$
\left<{\rm qhf}\right| \bar\Psi^1(x)\Psi^1(x) \left| {\rm qhf}\right> =\frac{1}{2}\sum_{m=0}^\infty\psi_m^\dagger(x)\psi_m(x)
$$
which differs from the result above by a factor of one half.  Further, it agrees with the result that one would obtain
using the commutator normal ordered operator that we defined in (\ref{sigma}) using either route,
  $$
 \left<{\rm qhf}\right| \frac{1}{2}\left[\bar\Psi^{1 }(x),\Psi^1(x)\right] \left| {\rm qhf}\right>=\frac{1}{2}\sum_{m=0}^\infty\psi_m^\dagger(x)\psi_m(x) = \frac{B}{4\pi}
$$
This turns out to have the correct symmetry properties when combined with the other components
computed using the same technique.

There is another argument for the non-renormalization of the charge density, related to the fact that
an induced Chern-Simons term is also not renormalized beyond one loop.  The latter statement applies only when
there is a charge gap in the spectrum (as explicit computations show\cite{semenoffsodanowu}).  To be concrete, consider the Lagrangian of Coulomb
interacting graphene,
$$
L=\bar\psi i\gamma^\mu\partial_\mu\psi - \frac{e^2}{8\pi\epsilon}\psi^\dagger\psi\frac{1}{\sqrt{-\vec\nabla^2}}\psi^\dagger\psi
$$
which, by introducing a field $A_0$ can be written as the equivalent theory
$$
L=\bar\psi (i\gamma^\mu\partial_\mu+\gamma^0A_0)\psi + \frac{2\pi\epsilon}{e^2}A_0\sqrt{-\vec\nabla^2}A_0
$$
If we introduce gauge fields in the
central torus of U(4), $A^{\sigma\sigma'}={\rm diag}(A^1_\mu,A^2_\mu,A^3_\mu,A^4_\mu)$, and a C and P invariant
mass term $m^{\sigma\sigma'}={\rm diag}(m,m,-m.-m)$,
$$
L=\bar\psi (i\gamma^\mu\partial_\mu+\gamma^\mu A_\mu-m)\psi   + \frac{\pi\epsilon}{8e^2}(\sum_\sigma A^\sigma_0)\sqrt{-\vec\nabla^2}(\sum_\sigma A^\sigma_0)
$$
where we have written the last term so that the Coulomb interaction is mediated by the U(1) component of the
gauge field.

At one-loop order, integrating the fermions out of
the theory induces a Chern-Simons term
$$
S_{\rm CS}=\frac{e^2}{8\pi}{\rm sign}(m) \int d^3x\left(A^1dA^1+A^2dA^2-A^3dA^3-A^4dA^4\right)
$$
in the effective action.
By a straightforward generalization of the arguments in Refs.~\cite{Coleman:1985zi}\cite{Semenoff:1988ep},
it can be shown that this Cherm-Simons term does not obtain corrections beyond one loop order.  Furthermore,
this one-loop order result, when used to derive the U(4) charge density, gives precisely the result that we
have found. 

Depending on the U(4) orientation of the magnetization, the quantum Hall ferromagnet can have several
manifestations with significantly different physical properties.  First of all, it can be a genuine ferromagnet
with the spins of states in the charge neutral point Landau level aligned with the same orientation.
Recalling that the zero modes of one of the valleys have support only on one of the sublattices, and the other
valley on the other sublattice, the ferromagnet has electrons occupying both sublattices with equal amplitudes.
This is akin to a Wigner lattice.   In the same manifold of ground states is the anti-ferromagnet where spin
up lives on one sublattice and spin down lives on the other sublattice.   Then, there are other states where the
electrons are paired to spin singlets, but only inhabit one valley and therefore one sublattice.   We emphasize that
there is no distinction between these states at the level of the U(4) invariant theory, they are all in the manifold
of degenerate ground states that we call the ``quantum Hall ferromagnet state''.  Interactions which have been ignored
in the U(4) invariant continuum theory, but which would certainly be present in real graphene, would be important for
distinguishing which these states actually forms the ground state.  The energy scale of these symmetry breaking
interactions is typically of the order of a few percent of the Coulomb energy.

\section{Appendix: Sum rule}

  The wavefunction is given by
\begin{equation}
u_{nm} (\vec x)=\frac{1}{\sqrt{2}} \frac{(\alpha^\dagger)^{n}}{\sqrt{n!}}
\frac{(\beta^\dagger)^m}{\sqrt{m!}}f_0 (\vec x)
\end{equation}
where
\begin{equation}
\alpha^\dagger = \frac{i}{\sqrt{2B}}\left[2\partial_x -\frac{B}{2}\bar x\right]~,~
\beta^\dagger =\frac{i}{\sqrt{2B}}\left[2\bar\partial_x  -\frac{B}{2}x \right]~,~
f_0 = \sqrt{\frac{B}{2\pi}} e^{-\frac{B}{4}\bar x x}
\end{equation}
and we use the complex plane coordinates $x=x_1+ix_2$, $y=y_1+iy_2$
and we shall assume that $B$ is
positive. Recall that $\alpha^\dagger$ and $\beta^\dagger$
commute with each other.  We write the kernel as

\begin{align}
\sum_{m =0}^\infty   u_{nm}(x)u_{nm}^*(y)
 =& \frac{B}{2\pi}\sum_{m =0}^\infty   \frac{1}{ n !m!}[(\alpha^\dagger)^{n }(\beta^\dagger)^m f_0(x)]
[(\alpha^\dagger)^{n }(\beta^\dagger)^m f_0(y)]^* \\
 =& \frac{B}{2\pi}\sum_{m =0}^\infty   \frac{1}{ n !m!}\left(\frac{B}{2}\right)^m[(\alpha^\dagger)^{n }x^m f_0(x)]
[(\alpha^\dagger)^{n }y^m f_0(y)]^* \\
 =& \frac{B}{2\pi}\sum_{m =0}^\infty  \frac{1}{ n !m!}\left(\frac{B}{2}\right)^m (\alpha_x^\dagger)^{n }(\alpha_y^\dagger)^{ n *}
 x^m \bar y^m e^{-\frac{B}{4}(\bar x x +\bar y y)}  \\
 =& \frac{B}{2\pi}   \frac{1}{ n ! }  (\alpha_x^\dagger)^{n }(\alpha_y^\dagger)^{ n *}
  e^{-\frac{B}{4}(\bar x x +\bar y y)+\frac{B}{2}\bar y x}\\
 =& \frac{B}{2\pi} \frac{1}{ n ! }
\left(\frac{1}{2B}\right)^{n } \left[2\partial_x -\frac{B}{2} \bar x\right]^{ n }\left[2\bar\partial_y -\frac{B}{2}y\right]^{n }
  e^{-\frac{B}{4}(\bar x -\bar y)(x-y)+\frac{B}{4}(\bar y x - \bar x y)} \\
 =& \frac{B}{2\pi} \frac{1}{ n ! }
\left(\frac{1}{2 }\right)^{n } \left[2\partial_x -\frac{B}{2} \bar x\right]^{ n }\left[ x -  y\right]^{n }
  e^{-\frac{B}{4}(\bar x -\bar y)(x-y)+\frac{B}{4}(\bar y x - \bar x y)}  \\
=& \frac{B}{2\pi}  \frac{1}{(n)! }
\left(\frac{1}{2 }\right)^{n}  e^{-\frac{B}{4}(\bar x -\bar y)(x-y)+\frac{B}{4}(\bar y x - \bar x y)} \left[2\partial_x - B(\bar x-\bar y)\right]^{n}\left[ x - y\right]^{n}  \\
=& \frac{B}{4\pi} e^{-\frac{B}{4}(\bar x -\bar y)(x-y)+\frac{B}{4}(\bar y x - \bar x y)}  \frac{1}{(n)! }
 \left[\partial_x - B(\bar x-\bar y)/2\right]^{n}\left[x -  y\right]^{n}\\
=& \frac{B}{2\pi} e^{-\frac{B}{4}(\bar x -\bar y)(x-y)+\frac{B}{4}(\bar y x - \bar x y)} L_n( B|x-y|^2/2)
\end{align}
where $L_n$ are Laguerre polynomials.

\vskip .5cm

This work is supported in part by the Natural Sciences and Engineering Research Council of Canada.  One of the authors acknowledges the hospitality of Nordita, the Aspen Center for Physics, the Kavli Institute for Theoretical Physics and
the Galileo Galilei Institute where parts of this work were completed.

\end{document}